\documentclass[3p,preprint]{elsarticle}

\usepackage{natbib}
\usepackage{amsmath}
\usepackage{amssymb}
\usepackage{epsfig}
\usepackage{subfig}
\usepackage{tikz,siunitx}
\usepackage{xcolor}

\newcommand{\solidline}{\protect\tikz[baseline]{\protect\draw[solid,line width=1pt](0.0mm,0.5ex)--(6.5mm,0.5ex)}}
\newcommand{\dashline}{\protect\tikz[baseline]{\protect\draw[dashed,line width=1pt](0.0mm,0.5ex)--(6.5mm,0.5ex)}}
\newcommand{\dotline}{\protect\tikz[baseline]{\protect\draw[dotted,line width=1pt](0.0mm,0.5ex)--(6.5mm,0.5ex)}}
\newcommand{\dotdashline}{\protect\tikz[baseline]{\protect\draw[dash dot,line width=1pt](0.0mm,0.5ex)--(6.5mm,0.5ex)}}

\definecolor{pgreen}{RGB}{0,127,0}
\definecolor{pmagenta}{RGB}{191,0,191}

\title{Numerical and modeling error assessment of large-eddy simulation using
direct-numerical-simulation-aided large-eddy simulation}
\author[1]{H. Jane Bae\corref{cor1}}
\ead{jbae@caltech.edu}
\author[2]{Adri\'an Lozano-Dur\'an}
\ead{adrianld@stanford.edu}

\cortext[cor1]{Corresponding author}
\address[1]{Graduate Aerospace Laboratories, California Institute of
Technology, Pasadena, California 91125, USA}
\address[2]{Department of Aeronautics and Astronautics, Massachusetts
Institute of Technology, Cambridge, MA 02139, USA}

\begin{document}

\begin{abstract}  
We study the numerical errors of large-eddy simulation (LES) in isotropic and wall-bounded turbulence. A direct-numerical-simulation (DNS)-aided LES formulation, where the subgrid-scale (SGS) term of the LES is computed by using filtered DNS data is introduced. We first verify that this formulation has zero error in the absence of commutation error between the filter and the differentiation operator of the numerical algorithm. This method allows the evaluation of the time evolution of numerical errors for various numerical schemes at grid resolutions relevant to LES. The analysis shows that the numerical errors are of the same order of magnitude as the modeling errors and often cancel each other. This supports the idea that supervised machine learning algorithms trained on filtered DNS data might not be suitable for robust SGS model development, as this approach disregards the existence of numerical errors in the system that accumulates over time. The assessment of errors in turbulent channel flow also identifies that numerical errors close to the wall dominate, which has implications for the development of wall models.
\end{abstract}

\maketitle

\section{Introduction \label{sec:intro}}

DNS of most turbulent flows relevant for industrial applications is not tractable because the range of scales of motions in these flows is so large that the computational cost becomes prohibitive. In LES, the effect of the small scales on the larger ones is modeled through an SGS model, reducing the computational cost by several orders of magnitude. The accuracy of the representation of turbulent flows depends on both the performance or modeling capabilities of the choice of the SGS model, the adequacy of the LES resolution, and the numerical method employed to discretize the equations. In addition, the various sources of error can interact, which may lead to further complications in the validation of SGS models in LES. Given that accurate predictions are required in many engineering and scientific applications, a close assessment of LES numerical errors is crucial.  This will also have implications for developing SGS models using supervised learning methods, which most often use filtered DNS data to train LES models without regarding the numerical errors.

Various SGS models have been introduced in the past decades such as
the eddy-viscosity models
\citep{Smagorinsky1963,Schumann1975,Kraichnan1976,Metais1992,Rozema2015},
stochastic models \citep{Yoshizawa1982,Leith1990}, similarity models
\citep{Clark1979,Leonard1997,Geurts1997}, and mixed models
\citep{Zang1993,Vreman1994}, among others. In addition, an important
development in SGS modeling arose with the dynamic procedure
\citep{Germano1991,Germano1992,Lilly1992}, which allows optimization
of parameters in the eddy-viscosity SGS models in accordance with the
turbulent flow that is simulated.  First works aiming to assess the
accuracy of SGS models include the pioneering investigation by
\citet{Clark1979}, who established the numerical study of decaying
isotropic turbulence as a reference benchmark, although the grid
resolutions and Reynolds numbers tested were highly constrained by the
computational resources of the time.  Since then, common benchmarks
for LES have broadened to include simple hydrodynamic cases such as
forced or decaying isotropic turbulence \citep{Metais1992}, rotating
homogeneous turbulence \citep{Kobayashi2001}, spatial or temporal
mixing layers \citep{Vreman1996, Vreman1997} and plane turbulent
channel flow
\citep{Piomelli1988,Germano1991,Chung2010,Lozano-Duran2019}, among
others. See \citet{Bonnet1998} for an overview of cases for LES
validation.

The validation cases mentioned above are representative of most LES
error quantification, where the numerical error and the modeling error
are not distinguished. \citet{Vreman1996b} was one of the first to
quantify the modeling and discretization errors separately for various
flow properties. In this analysis, modeling and numerical errors were
found to be of comparable magnitude and could partially cancel each
other. Further error analysis for LES was presented in
\citet{Geurts2002} in the context of the
``subgrid-activity'' parameter. More recently, \citet{Meyers2003}
differentiated the two errors and again showed that the partial
cancellation of both sources can lead to coincidental accurate
results. Along the same line, \citet{Meyers2007} studied the combined
effect of discretization and model errors, and a further series of
works resulted in the error-landscape-methodology framework reviewed
by  \citet{Meyers2011}, where it is stressed that the determination of
LES quality based on one single metric alone may provide misleading
results. However, in these works, the LES errors are quantified as
the error occurring in SGS models on a DNS grid by setting the filter
size of the SGS grid to the LES grid size. As most SGS models for LES
are not designed or expected to work in DNS grid resolutions, the
error defined above may not be an accurate depiction of the
true modeling error observed in LES grid resolutions. Moreover, this
method is able to assess SGS models when there is a clearly defined
filter size, which is not straightforward for anisotropic and
unstructured grids \citep{Trias2017}. Ideally, the modeling error
would compare a filtered DNS solution in an LES grid with the LES
solution. This can be made possible by first quantifying the correct 
numerical error for an LES using explicitly filtered LES.  

Previous works on explicitly filtered LES include the study of
\citet{Winckelmans2001}, who investigated a two-dimensional explicitly
filtered isotropic turbulence and channel flow LES to evaluate various
mixed subgrid/subfilter scale models. \citet{Stolz2001} implemented
the three-dimensional filtering schemes of \citet{Vasilyev1998} by
using an approximate deconvolution model for the convective terms in
the LES equations. \citet{Lund2003} applied two-dimensional explicit
filters to a channel flow and evaluated the performance of explicitly
filtered versus implicitly filtered LES. \citet{Gullbrand2003}
attempted the first grid-independent solution of the LES equations
with explicit filtering. \citet{Bose2010} further investigated the
grid independence of explicitly filtered LES with a three-dimensional
filter for turbulent channel flows. The recent works of
\citet{Bae2017,Bae2018b} introduce a DNS-aided LES (DAL) framework
using explicitly filtered LES that computes the exact SGS stress term
in the absence of commutation errors from filtered DNS data.

From a different perspective, machine-learning-based SGS models trained using high-fidelity data, typically DNS, have recently emerged (see reviews \citep{Kutz2017,Brenner2019,Duraisamy2019,Brunton2020}). SGS models have been developed in several canonical and complex flows, including two-dimensional homogeneous isotropic turbulence \citep{Maulik2019,Guan2022}, three-dimensional forced and \citep{Zhou2019,Xie2020,Frezat2021,Novati2021} decaying \citep{Wang2018,Beck2019} isotropic turbulence, turbulent channel flow \citep{Gamahara2017,Park2021,Bae2022}, and flow over an aircraft \citep{Lozano-Duran2020}. A majority of the research utilizes a supervised learning framework, where the model is trained to produce accurate SGS stresses obtained from DNS (or filtered DNS) data, and is often based on single-step target values to limit computational challenges in minimizing the model prediction error. Therefore, it is important to differentiate between \emph{a priori} and \emph{a posteriori} testing for these models. While the model might perform well in \emph{a priori} testing, \emph{a posteriori} testing is performed by integrating the LES equations in time. Due to the single-step cost function as well as the discrepancy in the filtered Navier-Stokes and LES equations, the resulting machine-learned model is not trained to compensate for the discrepancies between DNS and LES and the compounding (numerical and modeling) errors. The issue of inconsistency in data-driven SGS models has been exposed by studies that performed \emph{a posteriori} testing \citep{Nadiga2007,Gamahara2017,Beck2019}. Thus, in order to account for the discrepancies between DNS and LES in machine-learning-based model development, a systematic understanding of the different errors associated with LES SGS modeling is necessary.

It is the aim of this paper to introduce a framework that will allow
the evaluation of the time evolution of numerical errors
in LES. For that purpose, we introduce a method that produces zero
modeling error using the DAL framework \citep{Bae2017,Bae2018b} as a
method to define separate numerical and modeling errors. Our goal is
to assess the errors for unbounded and wall-bounded turbulent flows
using this methodology. In particular, for wall-bounded flows, the
wall-normal dependency of errors in LES will be investigated.

The paper is organized as follows. In Section \ref{sec:form}, the
filtered Navier-Stokes equations are compared to the traditional LES
equations to clearly define numerical error, and the DAL framework is
introduced. We then introduce the numerical experiments in Section
\ref{sec:errors}, where we evaluate the numerical errors for forced
isotropic turbulence and turbulent channel flows. Finally, we conclude
the paper in Section \ref{sec:conclusion}.

\section{Zero-modeling-error formulation \label{sec:form}}

\subsection{Filtered Navier-Stokes equations and large-eddy simulation
equations\label{sec:form:errors}}

The incompressible Navier-Stokes equations are given by
\begin{equation}\label{eq:NS_eq}
{\frac{\partial {u}_i}{\partial t}} + {\frac{\partial
{u}_i{u}_j}{\partial x_j}} = - \frac{1}{\rho}{\frac{\partial
{p}}{\partial x_i}}+\nu{\frac{\partial^2 {u}_i}{\partial x_j \partial
x_j}}, \quad 
{\frac{\partial{u}_i}{\partial x_i}} = 0,
\end{equation}
where the velocity components are represented by ${u}_i$ for $i =
1,2,3$ (or equivalently $u$, $v$, and $w$) for the directions $x_i$
($x$, $y$, and $z$), $\rho$ is the fluid density, $\nu$ is the
kinematic viscosity, and $p$ is the pressure. As LES is formally
derived from the filtered Navier-Stokes equations, we define a filter
operator on a variable $\phi$ in integral form 
\begin{equation}\label{eq:filter_def}
\bar \phi(\boldsymbol{x}) = 
\int_\Omega G(t,\boldsymbol{x},\boldsymbol{x'}) \phi(\boldsymbol{x'}) 
\mathrm{d}\boldsymbol{x'},
\end{equation}
where $\boldsymbol{x}=(x_1,x_2,x_3)$, $G$ is the filter kernel, and
$\Omega$ is the domain of integration. When Eq. (\ref{eq:NS_eq}) is
filtered with Eq. (\ref{eq:filter_def}), the resulting equations are
\begin{equation}\label{eq:NS_eq_f}
\overline{\frac{\partial {u}_i}{\partial t}} +
\overline{\frac{\partial {u}_i{u}_j}{\partial x_j}} = -
\frac{1}{\rho}\overline{\frac{\partial {p}}{\partial
x_i}}+\nu\overline{\frac{\partial^2 {u}_i}{\partial x_j \partial
x_j}}, \quad 
\overline{\frac{\partial{u}_i}{\partial x_i}} = 0.
\end{equation}
However, in LES, we have access to the filtered velocity quantities
$\bar{u}_i$ and not the full flow field $u_i$. In order to be able to
solve Eq.  \eqref{eq:NS_eq_f} for a practical LES applications, the
order of the filter and the differentiation operator must be inverted
such that the equations become 
\begin{equation}\label{eq:LES_eq}
{\frac{\partial\bar{u}_i}{\partial t}} + {\frac{\partial
\overline{\bar{u}_i\bar{u}_j}}{\partial x_j}} = - \frac{1}{\rho}
{\frac{\partial \bar{p}}{\partial x_i}}+\nu{\frac{\partial^2
\bar{u}_i}{\partial x_j \partial x_j}}-\frac{\partial\mathcal{T}_{ij}}
{\partial x_j}+\delta^\text{num}_i,
\quad
{\frac{\partial\bar{u}_i}{\partial x_i}}+\delta^\text{num} = 0,
\end{equation}
where $\delta^\text{num}_i$ and $\delta^\text{num}$ are the
commutation error due to the inversion of the order of operators. In
traditional LES, the commutation error is often neglected or assumed
to be zero regardless of the choice of filter and differentiation
operator.  However, in most numerical methods used for LES, these
commutation errors are not negligible even with the use of commutative
filters \citep{Marsden2002} as the commutation of the two operators is
guaranteed only up to a finite order.  Here, we retain
$\delta^\text{num}_i$ and $\delta^\text{num}$ to signify the
difference between the formulations in Eq. \eqref{eq:NS_eq_f} and Eq.
\eqref{eq:LES_eq}. The term $\mathcal{T}_{ij} = \overline{u_i u}_j
- \overline{\bar{u}_i\bar{u}_j}$ is the SGS stress tensor, which is
  modeled through an SGS model in LES. At each time step, the effect
of using an SGS model can be characterized as $\delta^\text{mod}_i =
\partial/\partial x_j \left(\mathcal{T}_{ij}^\text{SGS} -
\mathcal{T}_{ij}\right)$, where $\mathcal{T}_{ij}^\text{SGS}$ is the
modeled SGS stress tensor given by the choice of SGS model. In this
paper, we will categorize the time-integrated effect of
$\delta^\text{num}_i$ and $\delta^\text{num}$ as the numerical error,
$\mathcal{E}^\text{num}$ in the absence of $\delta_i^\text{mod}$.

\subsection{Zero-modeling error through DNS-aided LES
\label{sec:form:zero_error}}

%
\begin{figure}
\begin{center}
\includegraphics[width=0.8\textwidth]{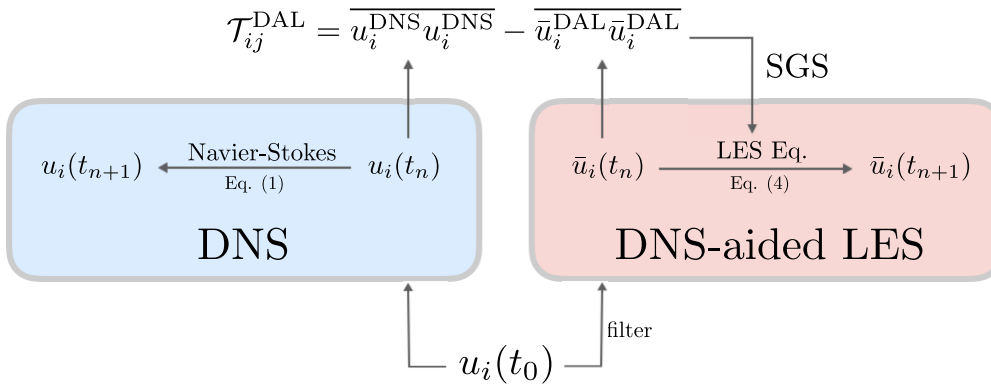}
\caption{Schematic of DAL. The SGS stress tensor
$\mathcal{T}_{ij}^\text{DAL}$ is computed at each time step from the
DNS flow field at the same time step and then applied to the LES
equations.  \label{fig:schematic}}
\end{center}
\end{figure}
One possible way of enforcing zero modeling error in LES is by using
an exact SGS model that can be produced by running a DNS concurrently.
That is, starting from an initial DNS velocity field
$u^\text{DNS}_i(t_0)$ and filtering it, we can compute the initial
condition $\bar{u}^\text{DAL}_i(t_0)$ for the DAL simulations such
that $\overline{u^\text{DNS}_i}(t_0) = \bar{u}^\text{DAL}_i(t_0)$. At
any given time step $t_n$ of the DAL, the exact SGS stress tensor
$\mathcal{T}_{ij}$ can be given by the DNS at the same time step
$t_n$, i.e.,
\begin{equation}\label{eq:DAL}
\mathcal{T}_{ij}^\text{DAL}(t_n) = 
\overline{u^\text{DNS}_i(t_n)u^\text{DNS}_j(t_n)} -
\overline{\bar{u}^\text{DAL}_i(t_n){\bar{u}^\text{DAL}}_j(t_n)}.
\end{equation}

A schematic of the proposed method is given in Figure
\ref{fig:schematic}. Note that this would be not possible for
traditional implicitly filtered LES due to the lack of an explicit
filter operator. In the absence of numerical (commutation) error, the
filtered DNS and DAL solutions will be identical, thus the name zero
modeling error. However, in the presence of numerical error, the
solutions of $\overline{u^\text{DNS}_i}$ and $\bar{u}_i^\text{DAL}$
will diverge from the integrated effect of $\delta_i^\text{num}$, and
the difference in the filtered DNS and DAL solution will provide a
measure for the numerical error of the simulation without the presence
of modeling error in a grid relevant to LES. 
%

\section{Error evaluation in LES \label{sec:errors}}

We consider two test cases, forced isotropic turbulence and turbulent
channel flow, to evaluate the numerical and modeling errors of LES.
We quantify the error in the LES flow fields as the $L_2$ error of the
filtered DNS and LES velocity fields,
\begin{align} \label{eq:mean_error}
\mathcal{E}^\text{tot}(t) =& \bigg[\frac{1}{L_xL_yL_z}\iiint
\sum_{i=1}^3\left(\overline{u_i^\text{DNS}}(t)
- \bar{u}_i^\text{LES}(t)\right)^2
\mathrm{d}x\,\mathrm{d}y\,\mathrm{d}z\bigg]^{1/2},
\end{align}
where $L_{x,y,z}$ is the length of the domain in the corresponding
direction. The error $\mathcal{E}^\text{tot}$ can be seen as a
combination of the numerical and modeling error by considering the
square-root of the integrand of Eq. \eqref{eq:mean_error} as
\begin{equation}
\overline{u_i^\text{DNS}}(t) - \bar{u}_i^\text{LES}(t)
=\left(\overline{u_i^\text{DNS}}(t) - \bar{u}_i^\text{DAL}(t)\right) 
+\Big(\bar{u}_i^\text{DAL}(t) - \bar{u}_i^\text{LES}(t)\Big), 
\end{equation}
where we define the numerical error as
\begin{equation} \label{eq:num_error}
\mathcal{E}^\text{num}(t) = \left[\frac{1}{L_xL_yL_z}\iiint
\sum_{i=1}^3\left(\overline{u_i^\text{DNS}}(t)-\bar{u}_i^\text{DAL}(t)\right)^2
\mathrm{d}x\,\mathrm{d}y\,\mathrm{d}z\right]^{1/2}.
\end{equation}
In essence, the numerical error quantifies the error due to the filter
operator.

In the case of channel flow, we also quantify a wall-normal distance
dependent error
\begin{equation} \label{eq:mean_error_y}
\mathcal{E}^\text{tot}_y(y,t) = \left[\frac{1}{L_xL_z}\iint
\sum_{i=1}^3\left(\overline{u_i^\text{DNS}}(y,t)
- \bar{u}_i^\text{LES}(y,t)\right)^2
\mathrm{d}x\,\mathrm{d}z\right]^{1/2},
\end{equation}
where $y$ is the wall-normal direction. An analogous $y$-dependent
numerical error can be defined as $\mathcal{E}^\text{num}_y$.

It is worth remarking the fact that if an SGS model is able to
counteract the numerical errors with the modeling errors such that
$\left(\overline{u_i^\text{DNS}}(t) - \bar{u}_i^\text{DAL}(t)\right)
=-\Big(\bar{u}_i^\text{DAL}(t) - \bar{u}_i^\text{LES}(t)\Big)$, the
total error is zero. Thus, not only does the DAL formulation provide a
method to evaluate errors, it can be used to inform future SGS models.
However, SGS models are not expected to provide the correct closure
term in an instantaneous manner. Rather, the goal of LES is to obtain
the correct statistics of the system. Thus, computing instantaneous
modeling errors does not align with the goal of LES and SGS modeling
and will not be considered in this paper.

\subsection{Forced isotropic turbulence \label{sec:errors:HIT}}

\begin{figure}
\begin{center}
\vspace{0.5cm}
\subfloat[]{\includegraphics[width=0.48\textwidth]{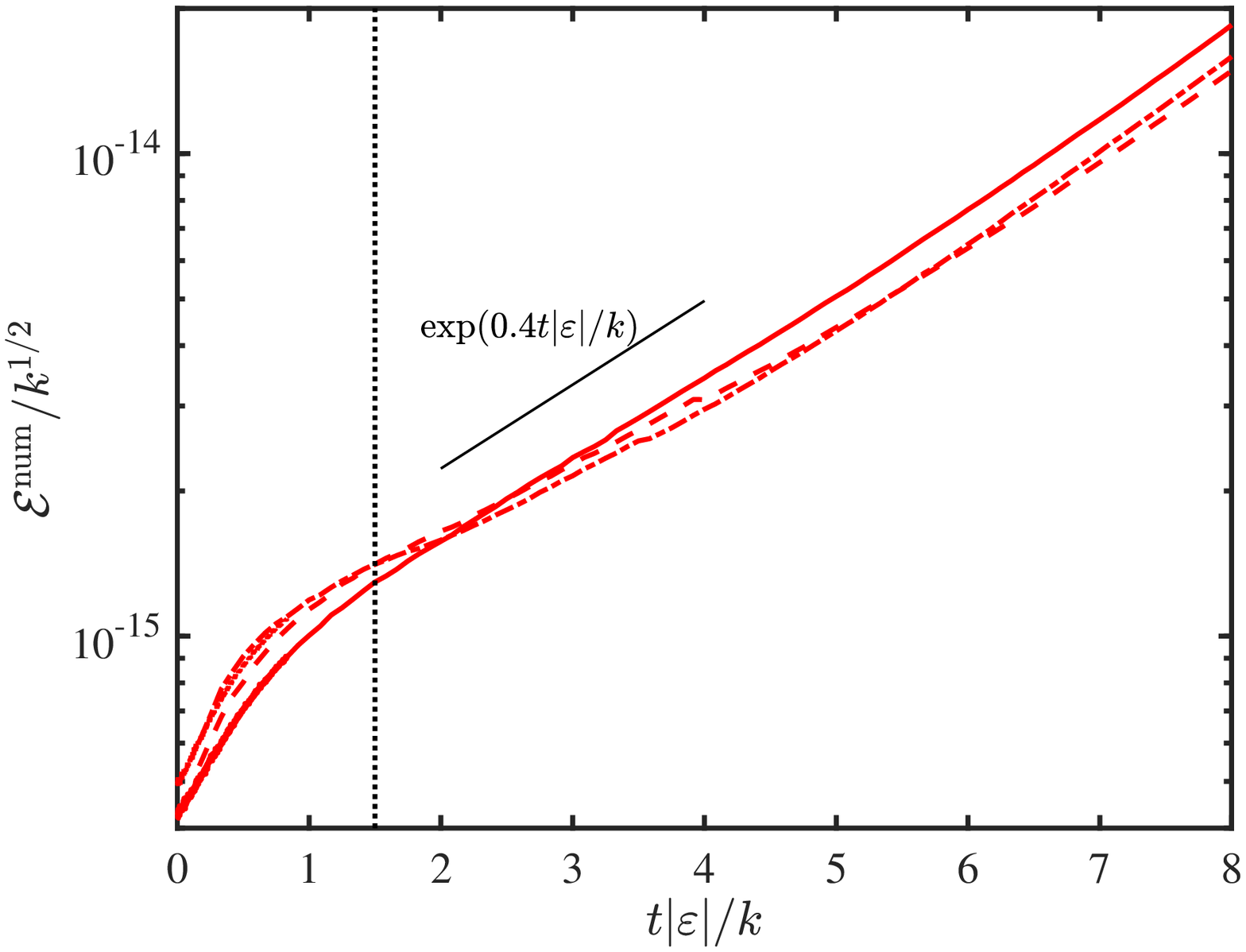}}
\hspace{0.2cm}
\subfloat[]{\includegraphics[width=0.48\textwidth]{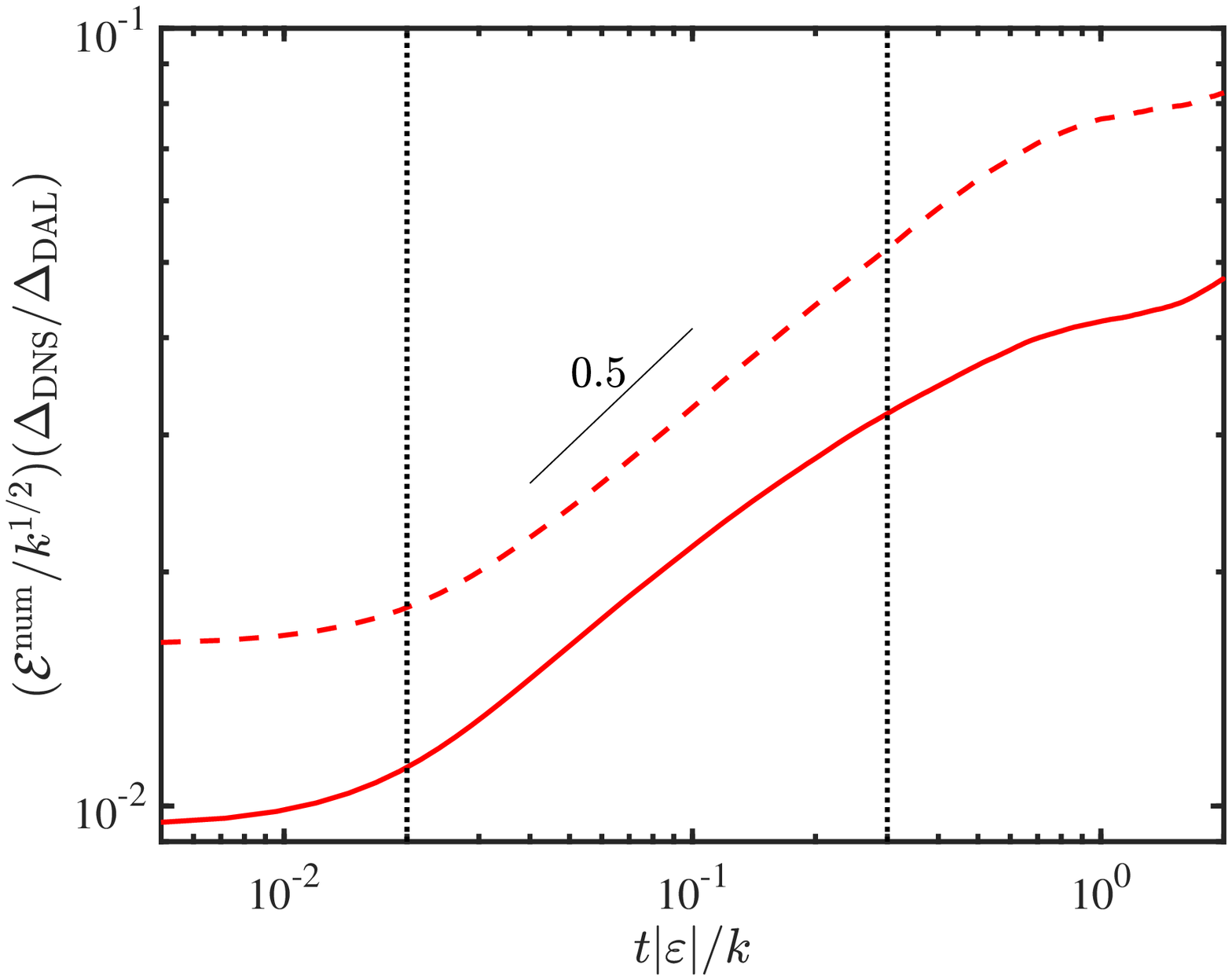}}
\caption{(a) Time evolution of the numerical error
$\mathcal{E}^\text{num}$ using dealiased Fourier discretization in
space and the Fourier cutoff filter for grid resolutions $32^3$
(\solidline), $64^3$ (\dashline), and $128^3$ (\dotdashline). Vertical
lines (\dotline) divide the initial response region and growth region.
(b) Time evolution of the numerical error $\mathcal{E}^\text{num}$
using a second-order finite difference in space and downsampling for grid resolutions $80^3$ (\solidline) and $48^3$ times (\dashline). Vertical lines (\dotline) divide the initial response region, growth
region, and saturation region.  
\label{fig:HIT_error}}
\end{center}
\end{figure}

To demonstrate the method given in section \ref{sec:form:zero_error}
provides zero modeling error and evaluate modeling error, we first
perform a simulation of forced isotropic turbulence using a dealiased
Fourier discretization in space in a triply periodic domain of size
$2\pi^3$ and advanced in time using a fourth-order Runge-Kutta time
stepping method. A linear forcing \citep{Lundgren2003}, $f_i = A u_i$,
is applied with $A = 0.45$. The DNS is performed using a $256^3$ mesh
at $Re_\lambda \approx 100$.  Three DAL (section
\ref{sec:form:zero_error}) cases with grid resolution of $32^2$,
$64^3$ and $128^3$ are performed with the filter operator given by a
sharp Fourier cutoff filter at the largest wavenumber resolvable by
the LES grid. The LES simulations are run with the same time-step size
as DNS to take advantage of the DAL formulation without interpolation
in time. The initial condition for the numerical experiment was taken
from a separate DNS after transients. The errors are normalized by
the square-root of the turbulence kinetic energy $k$, and the time is
given in integral times $k/|\varepsilon|$, where $\varepsilon$ is the
dissipation rate.

We evaluate the numerical error using DAL. Since DAL can be thought of
as supplying the exact SGS model that produces zero modeling error,
the difference between the filtered DNS flow field and the DAL flow
field can be considered the numerical error as defined in the beginning
of section \ref{sec:form:errors}. The sharp Fourier cutoff filter used
in the simulations commutes with the derivative operator in Fourier
space, which gives, in theory, zero numerical error. In practice, the
numerical error is zero only up to machine precision, and the
resulting small perturbations in the velocity profile will grow
exponentially following the leading Lyapunov exponent of the
Navier-Stokes equation until it reaches saturation \citep{Nastac2017}.
The Lyapunov exponent $\lambda$ represents the rate of separation,
such that the error grows proportional to $\exp(\lambda t)$, and its
reciprocal is closely related to the predictability horizon of a
chaotic solution, such as solutions to the Navier-Stokes equations.
Thus, DAL solution should give $\mathcal{E}^\text{num} \approx 0$
initially with a small exponential growth rate to ensure that this
formulation is effective.

In Figure \ref{fig:HIT_error}(a), we plot $\mathcal{E}^\text{num}(t)$
for the cases considered. As expected, the error is initially
$O(10^{-15})$ for all LES grid resolutions, with the error following a
linear curve, known as the initial response, until
$t|\varepsilon|/k\approx 1.5$.  Then, the error grows exponentially in
time with an exponential growth rate, or the Lyapunov exponent, of
$\lambda k/|\varepsilon| \approx 0.4$. This demonstration shows that
for short times, the two flow fields can be considered identical (the
error will be less than $10^{-10}$ for 30 integral times), and thus,
the modeling error evaluated from DAL flow fields is  zero in the
absence of numerical error. 

We repeat the previous experiment using a second-order finite
difference staggered discretization in space and advanced in time
using a third-order Runge-Kutta time-stepping method. The same linear
forcing is applied using a $240^3$ mesh. The DAL cases are three
and five times coarser than the DNS case in all three spatial
directions, i.e., computed on a $80^3$ and $48^3$  mesh. Due to the
numerical method of the simulation, there is no filter operator that
maps a velocity field of a DNS grid to a smaller LES grid and commutes
with the differentiation matrix (see Appendix A). An example of a case
with the LES grid equal to the DNS grid is given in \citet{Bae2018b}
for a differential filter \citep{Germano1986} utilizing the extension
method \citep{Bae2017,Bae2018b}. However, for practical applications,
LES should reduce the computational cost by using a low pass filter
and a smaller computational domain. Here, we utilize a down-sampling
method where the velocity field at a certain point of the LES is the
same as the (interpolated) DNS velocity field at the same point. 

The numerical error $\mathcal{E}^\text{num}$ as a function of time is
given in figure \ref{fig:HIT_error}(b). Unlike the case using Fourier
discretization where the differentiation operator commutes with the
Fourier-cutoff filter, the down-sampling does not commute with the
second-order finite-difference operation. The error is much larger
from the first time step, assuming a polynomial growth with a growth
factor of $0.4$ until it reaches saturation. Unlike the previous case,
the initial error is too large and the gap between the initial and
fully saturated error is too small to develop an exponential growth
region. Instead, the polynomial growth rate shows how the initial
error develops.  

\subsection{Channel flow \label{sec:errors:channel}} 

We perform a set of plane turbulent channel simulations at friction
Reynolds number $Re_\tau=u_\tau h/\nu\approx180$ and $550$, where $h$
is the channel half-height and $u_\tau$ is the friction velocity at
the wall. The simulations are computed with a staggered second-order
finite difference \citep{Orlandi2000} and a fractional-step method
\citep{Kim1985} with a third-order Runge-Kutta time-advancing scheme
\citep{Wray1990}. The code has been validated in previous studies in
turbulent channel flows \citep{Lozano-Duran2016,Bae2018a,Bae2019}. The
size of the channel is $2\pi h \times 2 h\times \pi h$ in the $x$, $y$
and $z$ directions, which are the streamwise, wall-normal, and
spanwise directions, respectively. It has been shown that this domain
size is sufficient to accurately predict one-point statistics for
$Re_\tau$ up to 4200 \citep{Lozano-Duran2014}. Periodic boundary
conditions are imposed in the streamwise and spanwise directions, and
the flow is driven by imposing a constant mean pressure gradient. 

The DNS grid resolution utilizes $192\times 120\times 96$ grid points
for the $Re_\tau\approx 180$ case and $510\times 300\times 510$ grid
points for the $Re_\tau\approx 550$ case in the three spatial
directions.  The streamwise and spanwise directions are uniform with
$\Delta_x^+\approx 6$ and $\Delta_z^+\approx 6$ or $3$, where the
superscript $+$ denotes wall units given by $\nu$ and $u_\tau$.
Non-uniform meshes are used in the normal direction, with the grid
stretched toward the wall according to a hyperbolic tangent
distribution. The height of the first grid cell at the wall is
$\Delta_y^+ \approx 0.2$ for both cases. The DAL cases are
either three or five times coarser than the DNS case in all three
spatial directions. 
Similar to the forced isotropic turbulence case, we
utilize a down-sampling method where the velocity field at a certain
point of the LES is the same as the (interpolated) DNS velocity field
at the same point. 
The details of the simulations are given in Table
\ref{tab:channel_cases}.
\begin{table} 
\begin{center} 
\setlength{\tabcolsep}{12pt}
\begin{tabular}{l l c c c c c c} 
Case       & Method & $Re_\tau$           & $\Delta_x^+$ & 
$\min(\Delta_y^+)$  & $\max(\Delta_y^+)$  & $\Delta_z^+$ \\ 
\hline
\hline
CH180DNS   & DNS    & $186$               & 6.1          &
0.19                & 8.2                 & 6.1          \\
CH180DAL3  & DAL    & $186$               & 18.3         &
0.61                & 24.3                & 18.3         \\
\hline
CH550DNS   & DNS    & $547$               & 6.7          &
0.21                & 9.6                 & 3.4          \\
CH550DAL3  & DAL    & $547$               & 20.2         &
0.66                & 28.7                & 10.1         \\
CH550DAL5  & DAL    & $547$               & 33.6         &
0.97                & 49.5                & 16.8         \\
\hline 
CH550DSM3  & DSM    & $547$               & 20.2         &
0.66       & 28.7                & 10.1         \\
CH550AMD3  & AMD    & $547$               & 20.2         &
0.66       & 28.7                & 10.1         \\
CH550NM3   & NM     & $547$               & 20.2         &
0.66       & 28.7                & 10.1         \\
\hline
\end{tabular} 
\end{center}
\caption{Tabulated list of cases for the channel flow case. Cases are
named as CH[$Re_\tau$][Method][Coarsening Factor], where method is
either DNS or DAL and coarsening factor is the ratio of grid points
from DNS to DAL in one spatial direction.  
\label{tab:channel_cases}}
\end{table}

\begin{figure}
\begin{center}
\vspace{0.5cm}
\subfloat[]{\includegraphics[width=0.48\textwidth]{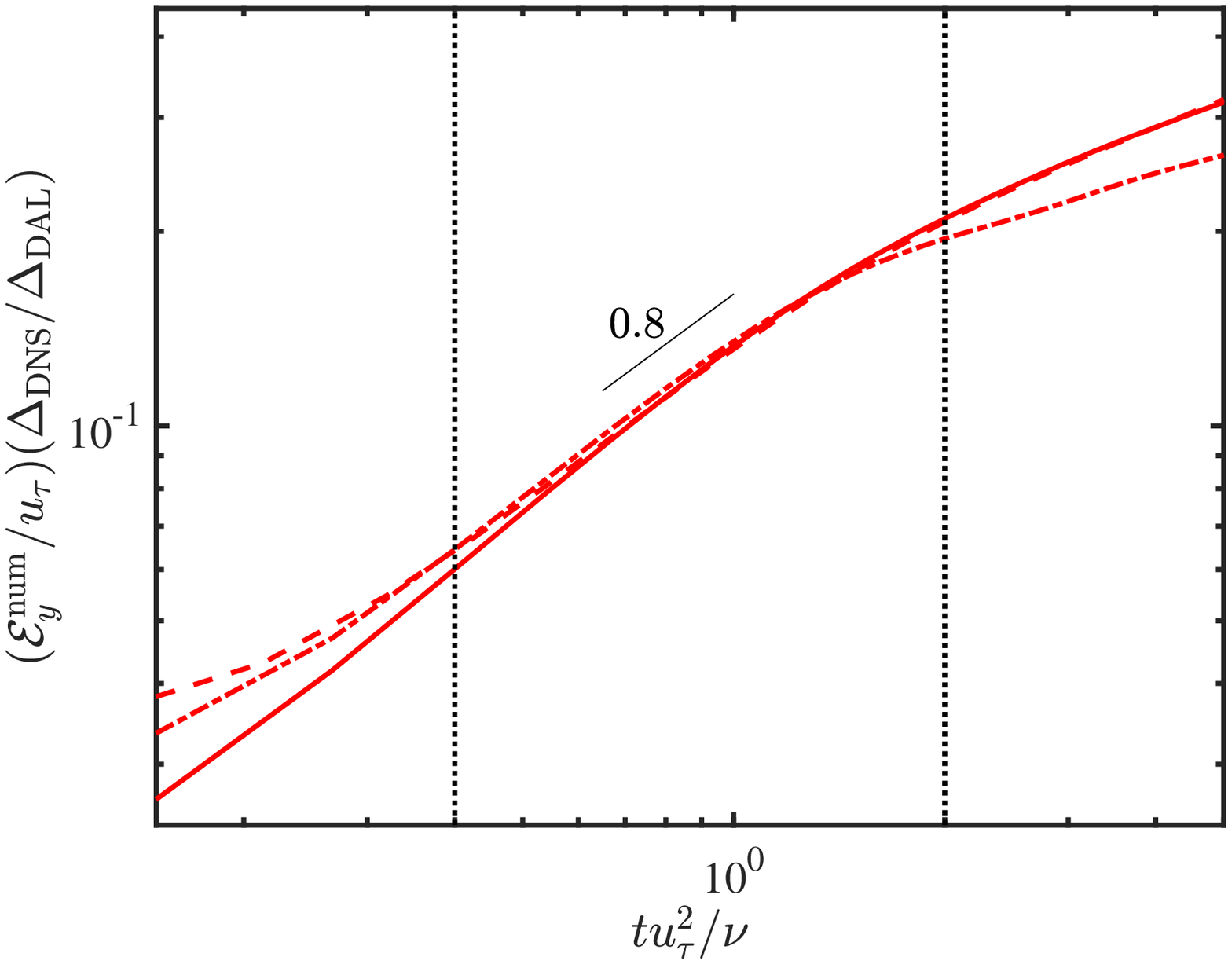}}
\hspace{0.2cm}
\subfloat[]{\includegraphics[width=0.48\textwidth]{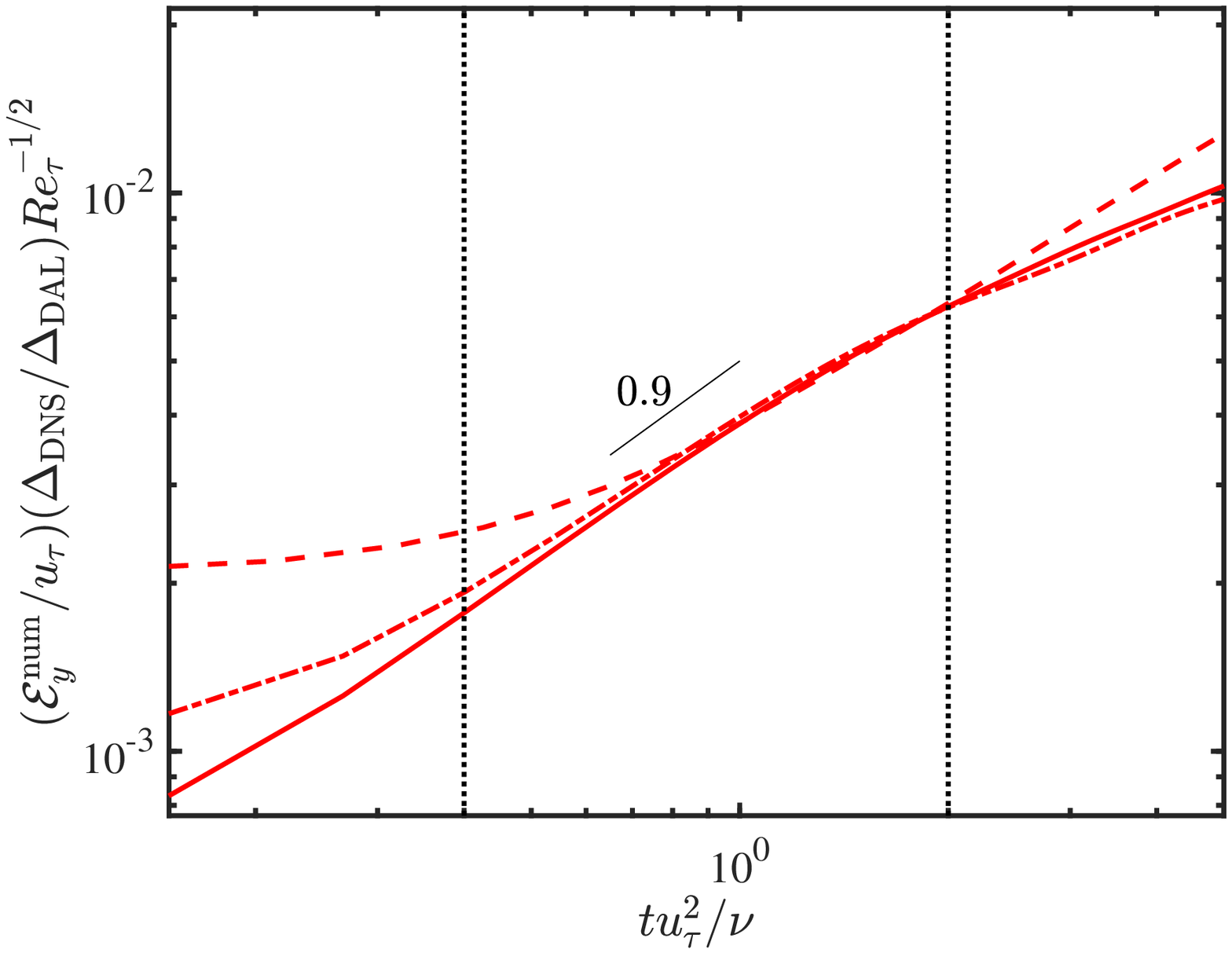}}\\
\subfloat[]{\includegraphics[width=0.48\textwidth]{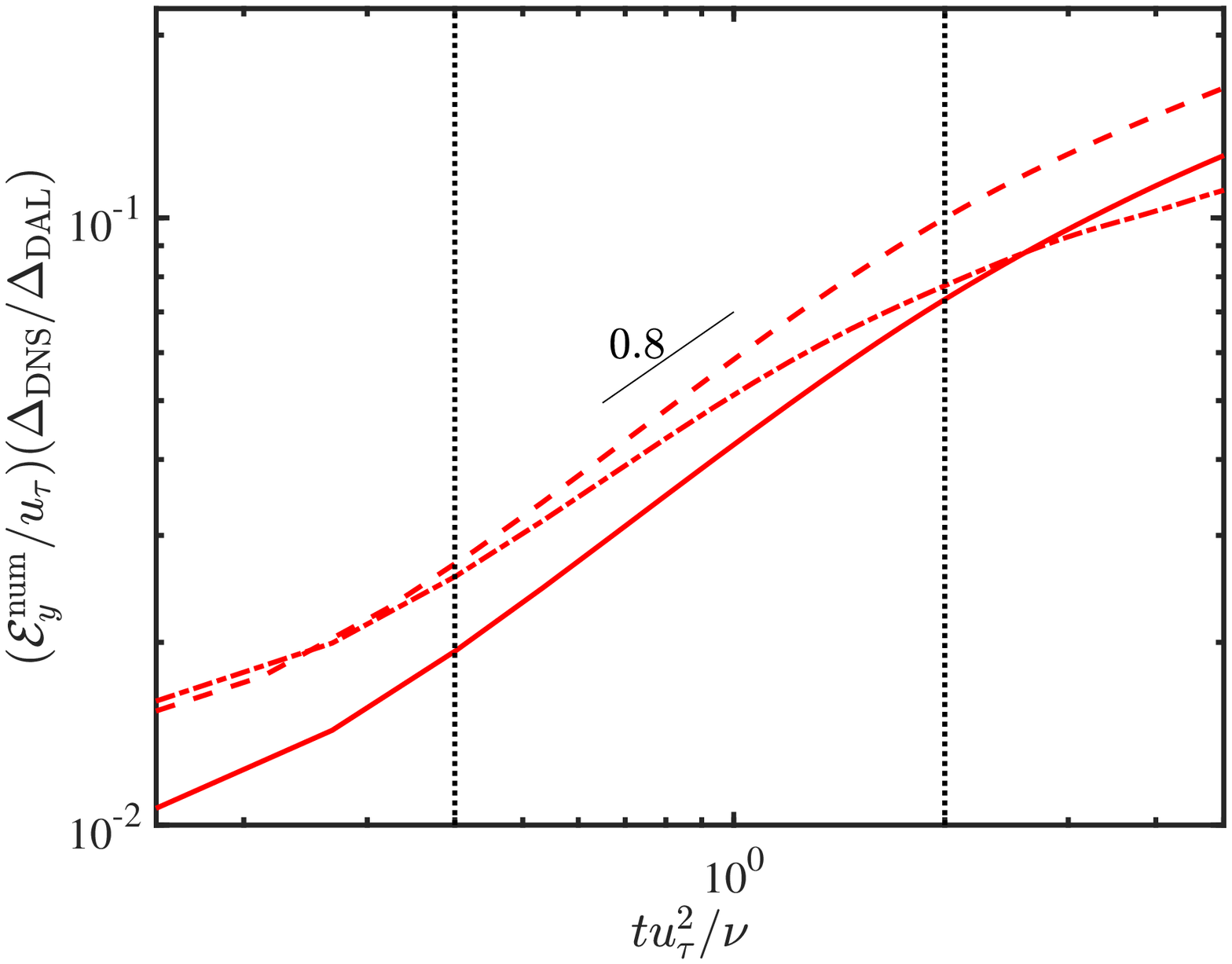}}
\hspace{0.2cm}
\subfloat[]{\includegraphics[width=0.48\textwidth]{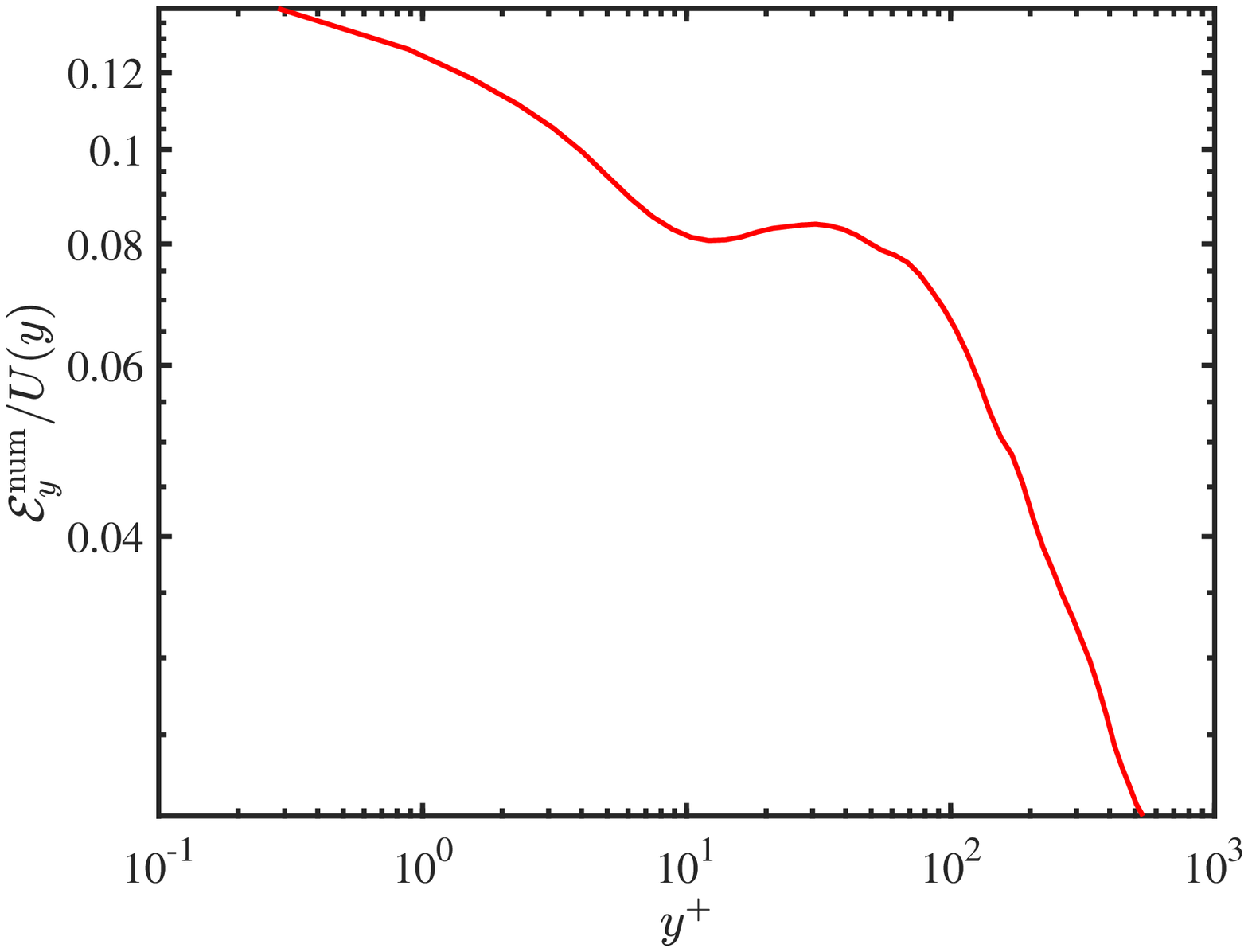}}
\caption{Time evolution of the (a) integrated numerical error
$\mathcal{E}^\text{num}$ and wall-normal distance dependent numerical
error $\mathcal{E}^\text{num}_y$ at (b) $y^+ \approx 15$ and (c) $y/h
\approx 1$ for CH180DAL3 
({\color{red}\dashline}), CH550DAL3 
({\color{red}\solidline}), and CH550DAL5
({\color{red}\dotdashline}). Vertical lines (\dotline)
divide the initial response region, growth region, and saturation
region. (d) $\mathcal{E}^\text{num}_y$ normalized with the mean
velocity profile as a function of wall-normal 
height at $tu_\tau^2/\nu \approx 1$ for CH550DAL3. 
\label{fig:channel_num_error}}
\end{center}
\end{figure}
In order to evaluate the numerical error, we use the same approach as
in section \ref{sec:errors:HIT} and supply the SGS stress term using
DAL for the three different cases.  Figure
\ref{fig:channel_num_error}(a) shows the evolution of the integrated
numerical error as a function of time. When the error is normalized
with $\Delta_\text{DNS}/\Delta_\text{DAL}$, we see that the error
evolution of the three cases collapses. The linear dependence of the
error on grid resolution and its independence on Reynolds number is
similar to the LES error scalings found in \citet{Lozano-Duran2019}.
This study did not distinguish between modeling and numerical errors
and considered errors in resulting statistics. However, this shows
that either the numerical errors are dominant in the total LES error
or the modeling error is proportional to the numerical error,
demonstrating that numerical errors are a significant source of error
in LES.

For all three cases, after the initial linear growth, the errors
assume a polynomial growth with a growth rate of 0.8, until
saturation. This growth rate is larger than that observed in the
forced isotropic turbulence case. The faster growth in error can be
attributed to the non-uniform mesh in the wall-normal direction and,
incidentally, the modifications in the  Poisson solver, which
contribute towards additional sources of numerical error. 

Figures \ref{fig:channel_num_error}(b) and (c) show
$\mathcal{E}^\text{num}_y$ for a given time at different wall-normal
heights, $y^+\approx 15$ and $y/\delta = 1$.  The linear dependence on
the factor $\Delta_\text{DNS}/\Delta_\text{DAL}$ is true for all
wall-normal heights, similar to the integrated error.  The growth
rates are similar for the two wall-normal heights, with the error
accumulating slightly slower in the outer region, compared to the
error at $y^+\approx 15$. This shows that the numerical error will
accumulate faster closer to the wall.  Additionally, the errors are
weakly dependent on Reynolds number close to the wall, but become
independent of Reynolds number away from the wall. Thus, the relative
error close to the wall will increase with Reynolds number but stay
constant with Reynolds number in the outer region.  Moreover, the
relative $y$-dependent numerical error at each wall-normal height is
given in figure \ref{fig:channel_num_error}(d). While only one time
step is shown, all temporal locations show similar trends as a
function of wall-normal distance. This shows that the relative
numerical error is the largest close to the wall and decreases with
distance from the wall. This large and fast accumulation of errors in
the near-wall region is known to be problematic for SGS models and in
particular for wall models for LES. 

\begin{figure}
\begin{center}
\vspace{0.5cm}
\subfloat[]{\includegraphics[width=0.48\textwidth]{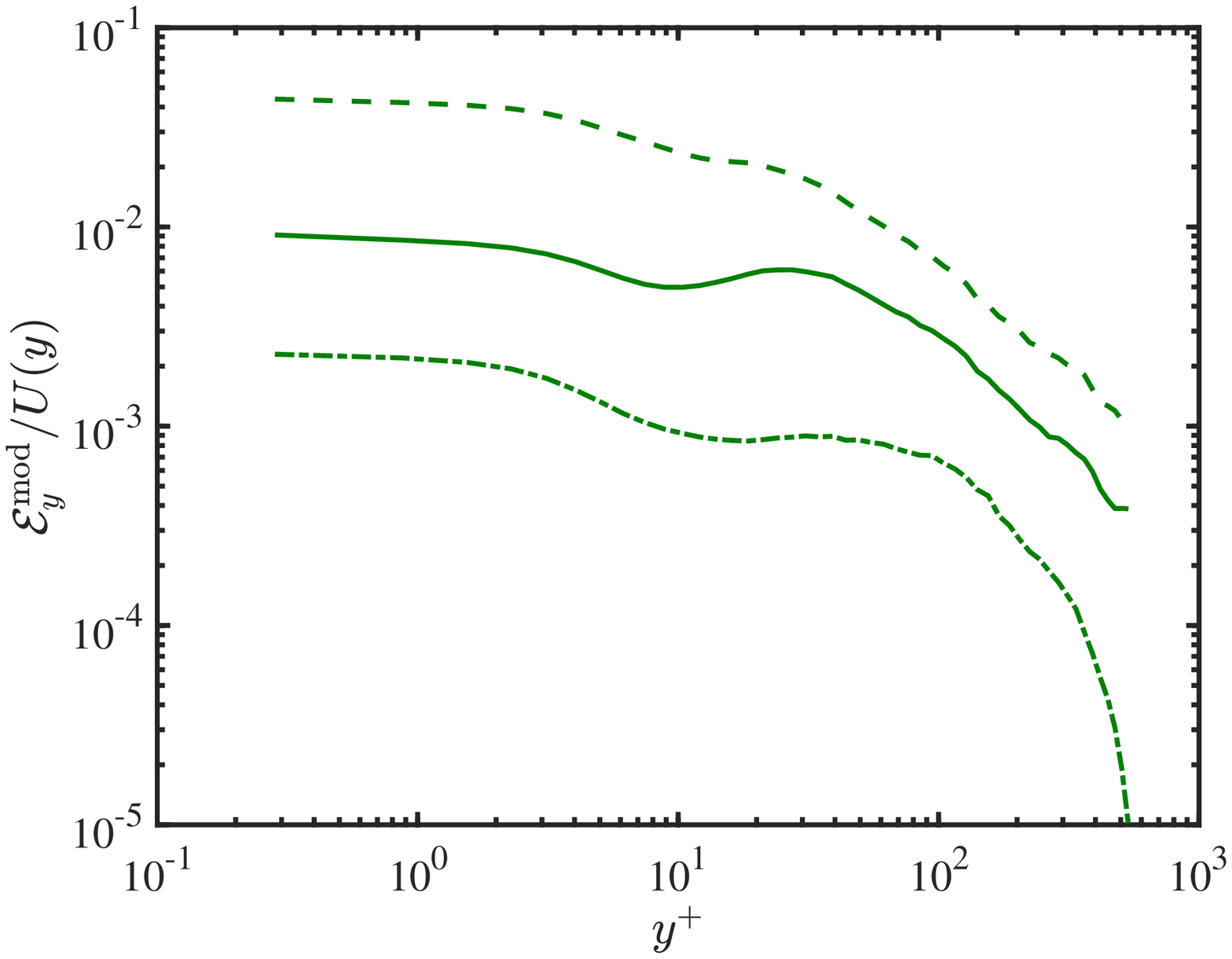}}
\hspace{0.2cm}
\subfloat[]{\includegraphics[width=0.48\textwidth]{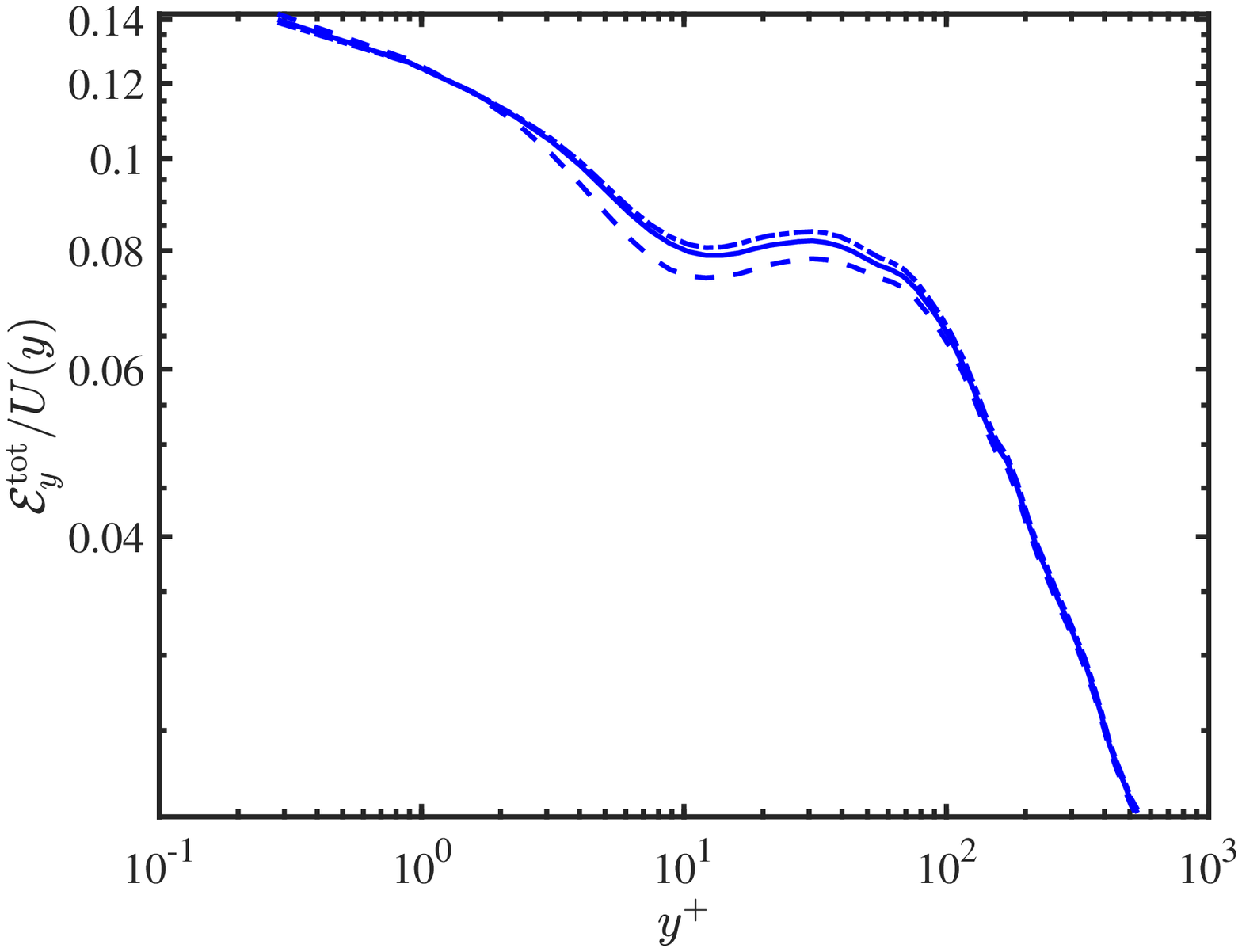}}
\caption{(a) $\mathcal{E}^\text{mod}_y$ and (b) $\mathcal{E}^\text{tot}_y$ normalized with the mean
velocity profile as a function of wall-normal 
height at $tu_\tau^2/\nu \approx 1$ for CH550DSM3 (\solidline), CH550AMD3 (\dashline), and CH550NM3 (\dotdashline). 
\label{fig:channel_mod_error}}
\end{center}
\end{figure}

Finally, we plot the modeling and total errors for the DSM and AMD model as well as the NM case in figure \ref{fig:channel_mod_error}. The errors show that the AMD model has the largest modeling error $\mathcal{E}_{mod}$, followed by the DSM and the NM case. However, the total error $\mathcal{E}_{tot}$ shows the smallest error for the AMD case, followed by the DSM and NM. This demonstrates that the modeling error and the numerical error cancel each other, as widely accepted in the literature. 

\section{Conclusions}\label{sec:conclusion}

Large-eddy simulation has become a viable tool in performing
simulations of turbulent flow for industrial and engineering
applications. In order for LES to be regarded as a useful tool, the
errors associated with it must be carefully studied. In particular,
the errors associated with the numerical method and the choice of
SGS model must be clearly distinguished and evaluated at
grid resolutions relevant to practical LES applications.  Moreover,
for wall-bounded flows, it is important to evaluate the error as a
function of wall-normal distance to account for near-wall effects.

In this paper, we introduce a formulation that allows the accurate
assessment of numerical errors at grid resolutions
typical of LES. DNS-aided LES is used to compute exact SGS stress
terms in the absence of numerical error; that is, by running a DNS and
LES with the equivalent initial condition, the DNS flow field at each
time step is used to generate the exact SGS stress term for the
corresponding LES flow field. This requires an explicit filter
operator which is often lacking in traditional LES. DAL allows the
evaluation of numerical errors as a function of time. Moreover, it
allows the distinction of numerical and modeling errors for LES,
which is often not straightforward. 

Using this method, we compute the numerical errors of two
cases: forced isotropic turbulence and turbulent channel flow. 
The evaluation of numerical errors of the Fourier discretization 
shows that the errors
are machine precision zero, as expected. This also verifies that DAL
indeed produces zero modeling error. In the case of 
numerical error for the finite difference case, the error follows a
polynomial growth until saturation. The growth rate is indicative of
how the LES solution is correlated to the corresponding DNS solution.
The results verify that the numerical errors are largest closer to the
wall. Furthermore, it also shows that numerical errors and modeling errors
partially counteract each other. The findings imply that SGS models
are not expected to perform well in all numerical methods and will
perform better when coupled with numerical methods where the numerical
errors are counterbalanced with the modeling errors. In particular, supervised learning SGS models trained on filtered DNS data do not account for numerical errors, and thus do not perform as expected in \emph{a posteriori} testing. The latter observation also has implications for wall modeling in LES, in which wall models become sensitive to the flow details close to the wall where the numerical errors are the largest.

The method introduced here can be helpful in two main directions.
Firstly, the method can inform the development of new SGS models in
conjunction with a  target numerical method such that the effect of
counterbalancing the numerical and modeling errors is maximized.
Since the method here provides the numerical errors as a function of
space and time, this information can be used to develop SGS models
specifically designed to cancel numerical errors. Secondly, the method
can similarly aid in the development and assessment of wall models, which
remains a pacing item to achieve practical LES.

\section*{Acknowledgments} 
H.J.B acknowledges support from the National Science Foundation under grant No.2152705.

\appendix
\section{Commutation of filter operator for finite difference}

For the filter operator to commute with the differentiation operator
numerically, we need, in matrix form,
\begin{equation} \label{eq:commute}
D^\text{LES} F\vec{u} = F D^\text{DNS}\vec{u},
\end{equation}
where a nonzero matrix $F\in\mathbb{R}^{N\times mN}$ is the filter
matrix, $D^\text{DNS}\in\mathbb{R}^{mN\times mN}$ and
$D^\text{LES}\in\mathbb{R}^{N\times N}$ are the one-dimensional
differentiation matrix for DNS and LES respectively. The factor $m$
denotes the ratio between the DNS and LES grid sizes, and for
practical cases, $m > 1$.  For simplicity, we consider the central finite
difference in a uniform grid, which gives
\begin{equation}
D^\text{DNS}_{i,j} = \begin{cases} 1/\Delta,& i = j+1  \\ -1/\Delta, &i =
j-1 \\ 0,&\text{otherwise}  \end{cases}, \quad
D^\text{LES}_{i,j} = \begin{cases} 1/m\Delta,& i = j+1  \\ -1/m\Delta, &i =
j-1 \\ 0,&\text{otherwise}  \end{cases},
\end{equation}
where $\Delta$ is the grid size for DNS.

Solving Eq. \eqref{eq:commute} as $F = D^{\text{LES}-1} F
D^\text{DNS}$, we get 
\begin{equation}
F_{i,j} = \begin{cases} 
\displaystyle{m\sum_{k = (i+1)/2}^{N/2} \left(-F_{2k,j-1} +
F_{2k,j+1}\right)},& i\text{ odd}\\
\displaystyle{m\sum_{k = 1/2}^{(i-1)/2} \left(F_{2k,j-1} -
F_{2k,j+1}\right)},& i\text{ even}
\end{cases}
\end{equation}
We can rewrite this equation as a linear equation $M\vec{f} = \vec{f}$
by transforming matrix $F$ as a vector $\vec{f}\in\mathbb{R}^{mN^2}$,
such that $\vec{f}_{mN(i-1)+j} = F_{i,j}$. Thus, there exists a filter
that commutes with the finite difference if and only if $M$ has an
eigenvalue of unity. While there may be a specific choice of $m$ and
$N$ such that this is possible (e.g. $m=1$), for an arbitrary grid,
this is not always true.  
 
\bibliographystyle{model1-num-names} 
\bibliography{eLES}

\end{document}